Nanoparticle doping as a way to enhance holmium fiber lasers efficiency

M. Kamrádek, I. Kašík, J. Aubrecht, P. Vařák, O. Podrazký, I. Bartoň, J. Pokorný, P. Peterka and P. Honzátko

Institute of Photonics and Electronics of the Czech Academy of Sciences, Chaberská 57, 182 51 Prague 8, Czech Republic

Abstract

Highly-doped holmium fibers have been prepared using modified chemical vapor deposition in combination with nanoparticle-doping method. Within a series of various $Al_2O_3$ and $Ho^{3+}$ concentrations, relations between fibers composition and their fluorescence and laser parameters have been studied. Al/Ho molar ratio equal to at least 50 was found to be the key factor for fibers with outstanding parameters. Fibers with slope efficiency above 80%, laser threshold below 100 mW and fluorescence lifetime up to 1.6 ms have been prepared. Thanks to high $Al_2O_3$ concentrations, obtained through nanoparticle doping, we were able to achieve high-performance fibers in a wide range of $Ho^{3+}$ concentrations. An output power of 19 W with 81% slope efficiency was reached using fiber with almost 4000 ppm of $Ho^{3+}$ and 10 mol.% of $Al_2O_3$. This result is encouraging for highly efficient high-power cladding-pumped holmium fiber lasers, and studied relations between fibers composition and their laser parameters will be used in the designing of such laser sources.

Introduction

Laser and optical fiber, two inventions awarded the Nobel Prize, coupled together form a fiber laser – a device which has shaped our lives in recent almost 40 years [1-4]. Compared to other types of solid-state lasers, fiber lasers offer advantages including high brightness, excellent beam quality, effective cooling, compactness and flexibility. Fiber lasers find utilization in various fields of human activity including communications, medicine, science, directed energy delivery or materials processing [5]. Without optical fibers and the world-famous erbium-doped fiber amplifier [6], invented by D. N. Payne and his team, the current form of the internet, and so our society, is unimaginable.

The heart of a fiber laser is formed by an active optical fiber doped with rare-earth (RE) ions; in this paper, we study $Ho^{3+}$ with emission around 2.1 μm. The most widespread matrix in fiber optics is silica glass, which excels in high optical transparency from ultraviolet to near-infrared, thermal durability, chemical stability and mechanical strength. On the other hand, its main drawbacks are high phonon energy and low miscibility with RE ions. To overcome them, the matrix is usually co-doped with aluminum oxide (alumina). Optical fibers are drawn from preforms prepared from extra pure precursors by advanced methods utilizing deposition from a vapor phase. Modified chemical vapor deposition (MCVD) method is the main technique used for the preform preparation [7, 8]. MCVD can be combined with solution doping, nanoparticle doping or with gas phase technique (chelate delivery system) to introduce desired dopants (alumina and $RE^{3+}$) into the preform core.

The solution-doping technique can be considered a standard or traditional method for incorporating dopants into an optical preform. This method has been developed by D. N. Payne and his colleagues in the 1980's [9]. The method is straightforward and versatile in terms of possible dopants. On the other hand, it is suitable mainly for moderately doped fibers. The practical limit of $Al_2O_3$ concentration is usually considered to be 5 mol.% of $Al_2O_3$ [10-13]. The core diameter is also limited because it is usually made from a single doped layer. Both drawbacks can be addressed by an advanced method using gas phase sublimated from solid precursors [14-16], either from halides or organometallic chelates. Commercially available devices include a heated cabinet and lines which need to be kept up to 250 °C to sublimate reactants and deliver them to the reaction zone. Utilizing gas-phase delivery, fibers containing up to 12 wt.% (~ 8 mol.%) of $Al_2O_3$ have been presented [17]. The nanoparticle-doping method might be a convenient alternative which enables high dopant concentrations and does not require any special, expensive and operationally demanding equipment. Fibers containing up to 9 mol.% of $Al_2O_3$ prepared using nanoparticles have already been presented [18, 19].

The motivation of using ceramic nanoparticles for active fibers and the history, lasting for about two decades, has already been reviewed elsewhere [20-23]. In 2019, we published an extensive comparison of holmium-doped fibers with various compositions up to 5 mol.% of $Al_2O_3$ prepared through solution doping and nanoparticle doping, and we did not observe any significant differences between both methods [21]. The best-performance fibers exhibited slope efficiency over 80% and fluorescence lifetime up to 1.35 ms, and the parameters were found to be dependent on the $Ho^{3+}$ concentration and the Al/Ho molar ratio not on the preparation method. In 2020 we figured out the key factor for highly-efficient fibers is not the $Ho^{3+}$ concentration itself but the Al/Ho ratio; the best-performance fibers need to have Al/Ho ratio at least 55 [24].

These results indicated the possibility of dissolution of $Al_2O_3$ nanoparticles or their reaction with silica matrix under high temperatures during the fibers processing. In 2021, we studied reactions between alumina nanoparticles and silica soot at high temperatures [25]. The $Al_2O_3$ nanoparticles reacted with $SiO_2$ at temperatures higher than 1600 °C to form mullite $3Al_2O_3·2SiO_2$, and no starting crystalline nanoparticles were found in the prepared fibers. All the results together suggest whether doping solution or dispersion is used the $Ho^{3+}$ vicinity in fibers is identical (or very similar at least). Regarding the concentrations of $Al_2O_3$ and $Ho^{3+}$ in the published papers, the fibers were rather lowly and moderately doped, i.e. generally $Al_2O_3$ 1–5 mol.% and $Ho^{3+}$ 400–3000 ppm. The presented work follows up on our previous results; we have moved the nanoparticle doping forward and scaled up the dopant concentrations.

The goal of this article is preparation of holmium-doped fibers optimized for fiber lasers with high concentrations of aluminum oxide and holmium ions while maintaining high slope efficiency, low laser threshold and long fluorescence lifetime. Fibers with various concentrations of $Al_2O_3$ up to 10 mol.% and $Ho^{3+}$ up to 6000 ppm have been prepared and tested in a laser arrangement. A special emphasis was placed on a high Al/Ho ratio to achieve a slope efficiency around 80%.

Experimental

Preparation of preforms and optical fibers

Optical preforms were prepared by the MCVD process under identical conditions. Fused silica tubes (F300, Heraeus) with OD/ID 18/15.2 mm were used as substrates. Extra pure $SiCl_4$ (Siridion STC 100, Evonik) and onsite purified oxygen (dew point <-99.9 °C) served as $SiO_2$ precursors. The deposition tube was at first polished at 2050 °C, and two glassy $SiO_2$ layers were deposited at 1780 °C as a buffer. After that, one porous core layer called a frit was deposited at 1300 °C. The frit was soaked with a doping suspension containing aluminum oxide nanoparticles (γ-$Al_2O_3$, 99.9%, Merck) and holmium (III) chloride ($HoCl_3·6H_2O$, 99.99%, Merck) in absolute ethanol. The frit was subsequently dried, sintered (under a chlorine atmosphere at temperatures 1200–1800 °C) and collapsed into an optical preform at temperatures up to 2100 °C. Optical fibers were drawn at 1950 °C as single mode ones with a cladding diameter of 125 μm (precision ± 1 μm). The fibers were coated with a UV-curable acrylate (Cablelite 3471-3-14, DSM Functional Materials) during drawing.

Preforms and fibers characterization methods

At first, refractive index profiles (RIPs) of the preforms were measured using an A2600 profiler (Photon Kinetics). The RIPs were determined at several positions along each preform, and under nine angles in each position to verify the preform homogeneity. Concentration profiles of $Al_2O_3$ were obtained by electron probe microanalysis (EPMA) using a JXA-8230 analyzer (Jeol). To obtain the profiles, polished preform cross-cuttings were analyzed in 20 equidistant points across the core.

The fibers RIPs were verified by an IFA-100 profiler (Interfiber Analysis). The fibers were next characterized according to their spectral absorption. Background losses were evaluated from absorption minimum (around 1300 nm) in a spectrum measured by a standard cutback method using an AQ6317B optical spectrum analyzer (Ando). Hydroxyl content was determined based on the absorption peak at 1383 nm. The concentration of $Ho^{3+}$ was calculated from the main absorption peak at 1950 nm measured with a Nicolet 8700 FT-IR spectrometer. The details of measurement and calculations can be found in [21].

Fluorescence lifetimes (FL) were measured using Innolume FBF-1150-PM-300 diode emitting at 1150 nm as an excitation source and a Hamamatsu G8371-01 InGaAs PIN photodiode as a detector. A short piece of fiber, approx. 2 mm, was used for the measurement. The decay curves were measured in a side detection setup; a detailed description of the setup can be found in [10, 19]. The decay curves were measured for multiple excitation powers, normalized and a decay time was obtained from 1/e value of intensity on the normalized curve. The fluorescence lifetime of the fiber was determined by extrapolation of the decay times to zero excitation power. The used methodology and setup serve to minimize the effect of various parasitic effects, such as reabsorption, amplified spontaneous emission and energy transfers [23].

Fiber laser setup

The fibers were tested according to a Fabry-Perot laser configuration depicted in Fig. 1. Tested fibers were pumped by a thulium-doped fiber (TDF), which was pumped by an erbium-doped fiber laser (EDFL). TDF laser cavity was formed using highly-reflective fiber Bragg grating (HRFBG) and lowly-reflective fiber Bragg grating (LRFBG) at 1950 nm at both ends of the fiber. The maximum output power at 1950 nm was about 1 W. The holmium-doped fiber laser cavity was formed by a HRFBG at 2100 nm (R~99.5%) and perpendicularly cleaved fiber end (R~3.5%). Tested fibers were gradually shortened to find the optimal fiber length. For each fiber length, laser threshold and slope efficiency (SE) with respect to the absorbed pump power were determined.

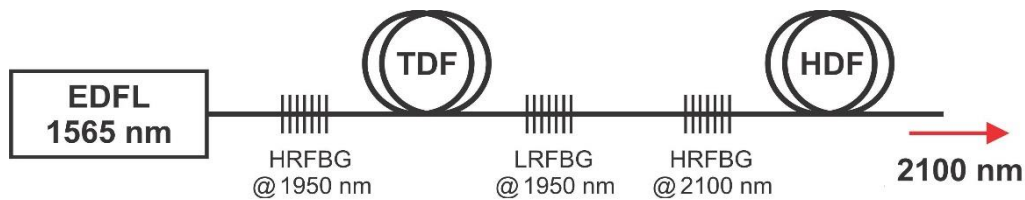

Fig. 1. Fiber laser setup.

Results and Discussion

The refractive index profiles of Fiber_9 and its preform are depicted in Fig. 2a). The preform profiles were taken in several z-positions along the preform to verify the preform homogeneity. The $Al_2O_3$ concentration profile in the preform of Fiber_9 can be found in Fig. 2b); the profile is in good agreement with the refractive index profiles. The relevant studied parameters of the fibers are listed in Table 1. In general, numerical aperture of the prepared fibers was in a range 0.24–0.27, the background losses varied 20–40 dB/km, and the fibers contained around 1 ppm of hydroxyl groups.

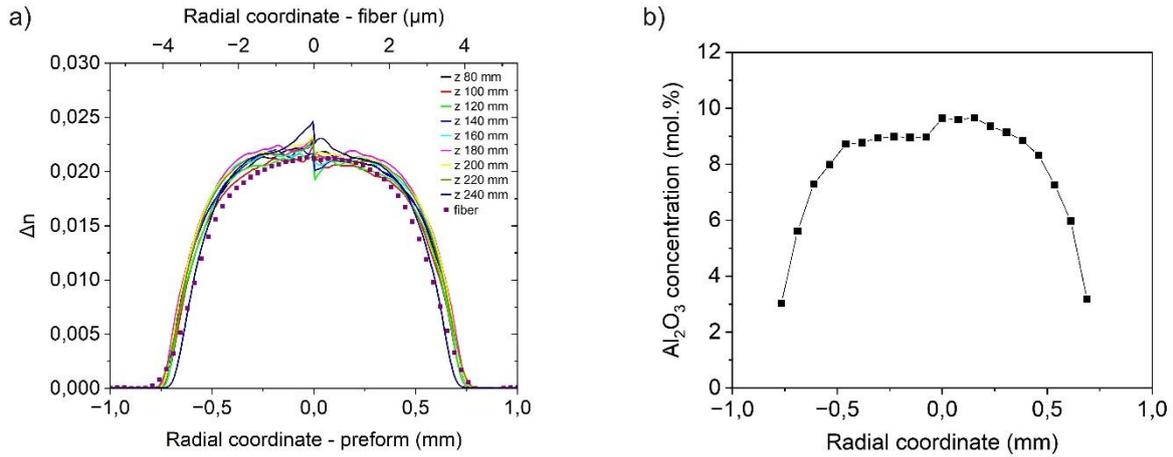

Fig. 2. a) Refractive index profiles of Fiber_9 and the preform, b) alumina concentration profile in preform of Fiber_9.

Table 1. Relevant parameters of fibers under study.

| Sample | $Al_2O_3$ (mol.%) | $Ho^{3+}$ (mol. ppm) | Al/Ho | Slope (%) | Threshold (mW) | Fluorescence lifetime (µs) | $L_{opt}$ (m) | $D_{core}$ (µm) |
|---|---|---|---|---|---|---|---|---|
| Fiber_1 | 10 | 2820 | 71 | 79.4 | 91 | 1488 | 1.0 | 7.4 |
| Fiber_2 | 8.8 | 6143 | 29 | 63.6 | 208 | 1190 | 0.5 | 8.1 |
| Fiber_3 | 8.8 | 5705 | 31 | 70.3 | 156 | 1200 | 0.6 | 7.4 |
| Fiber_4 | 10.4 | 1300 | 160 | 86.2 | 47 | 1610 | 2.0 | 6.0 |
| Fiber_5 | 10.3 | 2600 | 80 | 80.9 | 57 | 1470 | 1.4 | 6.0 |
| Fiber_6 | 10.6 | 2200 | 96 | 78.4 | 70 | 1510 | 1.0 | 6.9 |
| Fiber_7 | 10.3 | 1300 | 158 | 78.0 | 49 | 1560 | 1.8 | 7.1 |
| Fiber_8 | 9.8 | 1480 | 133 | 85.5 | 59 | 1460 | 1.4 | 7.3 |
| Fiber_9 | 9.1 | 2300 | 79 | 81.2 | 80 | 1391 | 1.0 | 7.2 |
| Fiber_10 | 9.3 | 2850 | 66 | 79.9 | 80 | 1350 | 0.7 | 7.2 |
| Fiber_11 | 9.4 | 3500 | 54 | 77.8 | 86 | 1312 | 0.7 | 7.1 |
| Fiber_12 | 9.4 | 3850 | 49 | 78.5 | 84 | 1247 | 0.6 | 6.5 |
| Fiber_13 | 9.3 | 1900 | 98 | 81.1 | 76 | 1370 | 1.3 | 6.3 |
| Fiber_14 | 10.0 | 5150 | 39 | 74.6 | 94 | 1260 | 0.5 | 6.7 |

The relations between fibers composition and the studied parameters are depicted in Figs. 3-5. Data acquired for lowly doped fibers already published in [21] are shown for comparison as well. The current data are denoted NP-high, the previously published are marked NP-low and SD for fibers prepared through nanoparticle and solution doping respectively.

Looking at SE as a function of $Ho^{3+}$ concentration, Fig. 3a, it is obvious, compared to the previously published data, we made a huge step forward. Thanks to high contents of $Al_2O_3$, we prepared high-efficiency fibers (75%) with $Ho^{3+}$ concentration above 5000 ppm. Even fibers with $Ho^{3+}$ around 6000 ppm showed SE above 60%. SE as a function of Al/Ho ratio is depicted in Fig. 3b, and it has been confirmed that this parameter is the key one for high efficiency. The data for high doping level are in a great agreement with the previous results. To obtain SE around 80%, the Al/Ho ratio needs to be at least 50.

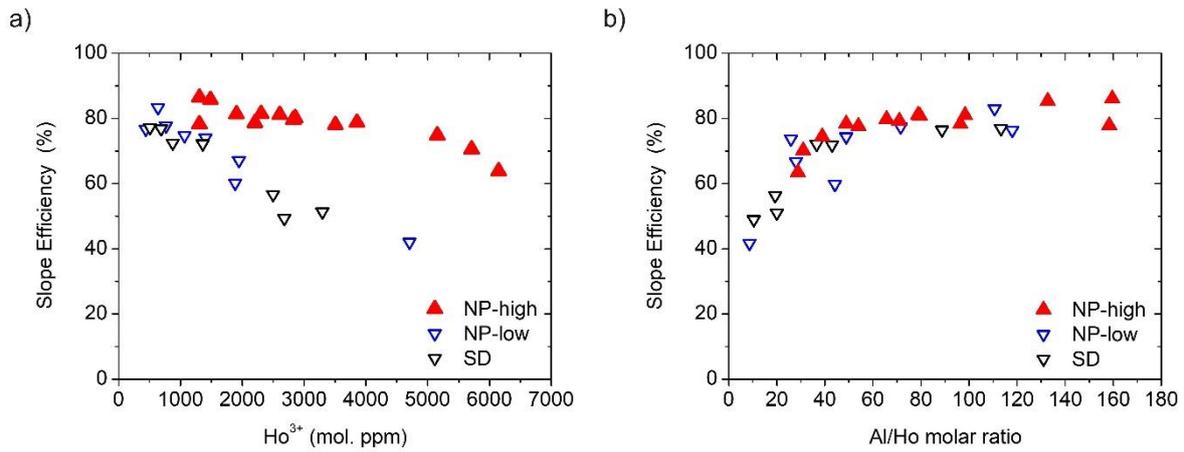

Fig. 3. Laser slope efficiency as a function of a) $Ho^{3+}$ concentration and b) Al/Ho molar ratio.

The dependencies of laser threshold can be found in Fig. 4. a) vs. $Ho^{3+}$ and b) vs. Al/Ho ratio. Compared to our previous results, the data are not so scattered, and the threshold values are much lower – in most cases below 100 mW. Similarly to SE, to obtain high-quality fibers, the Al/Ho ratio needs to be at least 50.

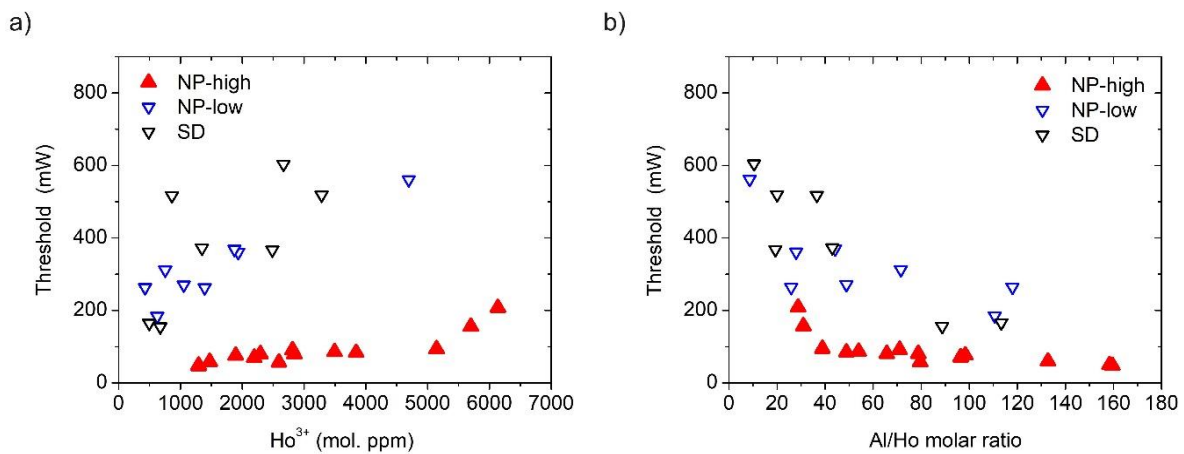

Fig. 4. Laser threshold as a function of a) $Ho^{3+}$ concentration and b) Al/Ho molar ratio.

The fluorescence lifetimes are depicted in Fig. 5 a) against $Ho^{3+}$ and b) against Al/Ho ratio. In general, long fluorescence lifetimes have been obtained and values in a range 1.2–1.6 ms confirm high quality of the fibers. To the best of our knowledge, fluorescence lifetime of 1.6 ms means a record value for silica-based holmium fibers.

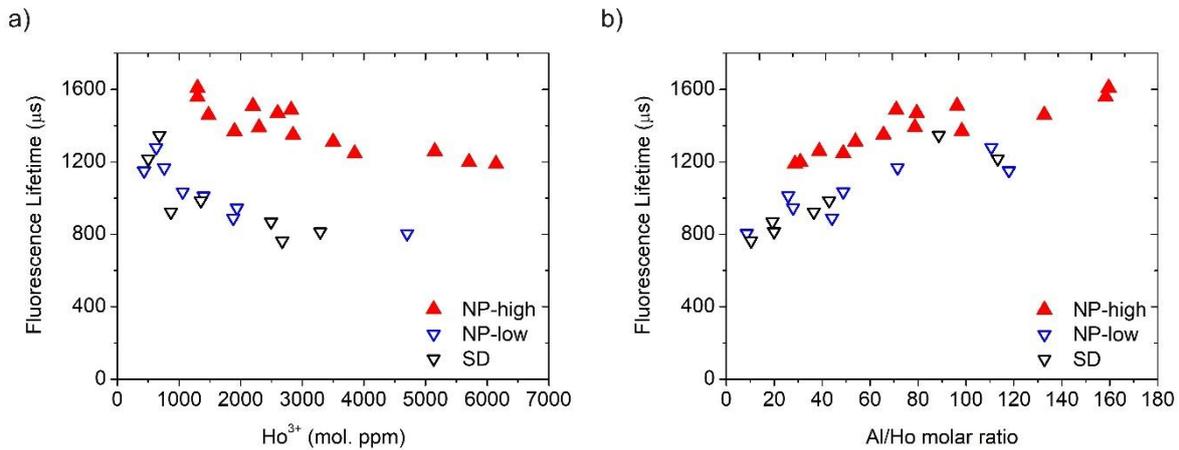

Fig. 5. Fluorescence lifetime as a function of a) $Ho^{3+}$ concentration and b) Al/Ho molar ratio.

Compared to SE and threshold, it seems the lifetime has no limitation, i.e. it raises in the whole tested range of Al/Ho ratio. In comparison with the previously published data, the lifetimes are higher even for similar Al/Ho ratios. These facts suggest $Ho^{3+}$ content has only a minor effect on fluorescence lifetime and the biggest influence comes from high content of $Al_2O_3$. Basically, alumina has two effects. First, it modifies silica matrix, and by creating non-bridging oxygens it compensates the positive charge of $Ho^{3+}$ ions and thus prevents their clustering and minimizes energy transfers between them. Second, alumina reduces the glass matrix phonon energy and thus decreases the probability of multi-phonon, non-radiative transitions [26-28]. The measured results for SE and threshold suggest, to minimize clustering, Al/Ho ratio around 50 is sufficient, and further increase has only a small effect. However, addition of $Al_2O_3$ (increasing Al/Ho ratio) further reduces the matrix phonon energy and thus improves the fluorescence lifetime.

The fluorescence lifetimes related to $Al_2O_3$ concentration have been studied in detail in thulium-doped fibers [29]. Similar lifetime improvement in highly-doped fibers has been observed. Moreover, a structural analysis of such fibers has been done, and phase-separated aluminum-enriched amorphous nanoparticles have been found. Although the $Tm^{3+}$ ions concentration was below the detection limit of the analysis, and no details about their distribution were presented, based on the relatively long fluorescence lifetimes up to 780 µs, the authors assumed that the active ions were embedded primarily within the aluminum-enriched nanoparticles, rather than in silica matrix where the lifetime would be about 400 µs.

Since the laser setup used for testing fibers was eligible only up to 1 W of pump power, selected highly-doped fibers, Fiber_12 and Fiber_14, were tested using a more powerful thulium laser with output power up to 25 W at 1939 nm. The setup in a Fabry-Perot arrangement consisted of thulium-doped fiber laser, fiber Bragg grating highly reflective at 2099 nm and tested holmium fiber perpendicularly cleaved to form a laser cavity. The tested fiber was placed on a cooling plate to control the fiber temperature. The results are listed in Table 2 together with a short review of remarkable state-of-the-art results obtained with core-pumped holmium-doped silica fiber lasers. From the table, it is obvious the prepared fibers, fabricated using nanoparticle-doping method, rank among the best-performance fibers published so far. Moreover, presented relations between fibers laser parameters and their composition offer a complex insight into the problematics and will be useful in the designing of high-power cladding-pumped fibers.

Table 2. Review of remarkable holmium-doped core-pumped silica fiber lasers.

| Reference | $Ho^{3+}$ conc. [mol. ppm] | Abs. @ 1950 nm [dB/m] | Output power [W] | Slope efficiency [%] |
|---|---|---|---|---|
| Pal 2016 [30] | 1500 | 43 | 7 | 74 |
| Hemming 2016 [31] | 2000 | 57 | 6 | 87 |
| Baker 2018 [32] | 4740 | 135 | 8 | 81 |
| Kamrádek 2020 [24] | 2320 | 66 | 6 | 81 |
| Beaumont 2023 [33] | 1510 | 43 | 30 | 86 |
| This work Fiber_12 | 3850 | 108 | 19 | 81 |
| Fiber_14 | 5150 | 145 | 16 | 76 |

Conclusions

We reported a series of 14 highly-doped holmium fibers with various compositions for a use in fiber lasers. The fibers were prepared using MCVD in combination with nanoparticle doping and contained in maximum around 10 mol.% of $Al_2O_3$ and about 6000 ppm of $Ho^{3+}$. We presented the relations between fibers composition and their laser and fluorescence parameters and found out that the key parameter for high-efficiency fibers is Al/Ho molar ratio greater than 50. Thanks to high contents of $Al_2O_3$, the prepared fibers showed slope efficiency around 80%, threshold below 100 mW and fluorescence lifetime longer than 1.2 ms in a wide $Ho^{3+}$ concentration ranges. The nanoparticle-doping method was proven to be suitable for preparation of highly-doped active fibers with excellent laser and fluorescence parameters.

Acknowledgements

This work was supported by by German Academic Exchange Service (DAAD-23-07). This work was supported by Czech Science Foundation (23-05701S). This work was co-funded by the European Union and the state budget of the Czech Republic under the project LasApp CZ.02.01.01/00/22_008/0004573.